# Realization of Surface-Obstructed Topological Insulators

Juan Du,[†] Tianzi Li,[†] Xiying Fan, Qicheng Zhang, and Chunyin Qiu[*]

Key Laboratory of Artificial Micro- and Nano-Structures of Ministry of Education and School of Physics and Technology, Wuhan University, Wuhan 430072, China

[†] These authors contributed equally.
[*] To whom correspondence should be addressed: cyqiu@whu.edu.cn

*Abstract.* Recently, higher-order topological insulators have been attracting extensive interest. Unlike the conventional topological insulators that demand bulk gap closings at transition points, the higher-order band topology can be changed without bulk closure and exhibits as an obstruction of higher-dimensional boundary states. Here, we report the first experimental realization of three-dimensional surface-obstructed topological insulators with using acoustic crystals. Our acoustic measurements demonstrate unambiguously the emergence of one-dimensional topological hinge states in the middle of the bulk and surface band gaps, as a direct manifestation of the higher-order band topology. Together with comparative measurements for the trivial and phase-transition-point insulators, our experimental data conclusively evidence the unique bulk-boundary physics for the surface-obstructed band topology. That is, the topological phase transition is determined by the closure of surface gap, rather than by closing the bulk gap. Our study might spur on new activities to deepen the understanding of such elusive topological phases.

*Introduction.*—The discovery of topological insulators (TIs) opens the door to various classes of topological phase of matter [1-3]. The band topology of a $d$-dimensional TI can be evidenced by the emergence of $(d-1)$-dimensional boundary states. This topological connection between the bulk and boundary is known as bulk-boundary correspondence [2,3]. Recently, a new family of topological phases, higher-order TIs (HOTIs), is proposed to support lower-dimensional boundary signatures [4-10], and thus go beyond the conventional bulk-boundary correspondence. Specifically, a $d$-dimensional $n$-order TI hosts $(d-n)$-dimensional boundary signatures (with $n > 1$). So far, despite few experiments reported in solid systems [11-13], the HOTIs are extensively realized in classical systems [14-38], such as mechanical [14,22], photonic [15,17-21], electric circuits [16-23], and acoustic [24-38] systems.

It is well known that a conventional TI cannot be adiabatically deformed to a trivial insulator without breaking the symmetry or closing the bulk energy gap, and the obstruction can be captured by topological invariants defined for bulk. The case becomes more subtle for HOTIs. Similar to the conventional TIs, some of the HOTIs (e.g. the breathing Kagome lattice model [10,24,25]) feature bulk closures at their topological transitions. Some of the HOTIs, now dubbed boundary-obstructed TIs (BOTIs), however, are distinguished from their trivial counterparts by closing higher-dimension boundary states [39-44]. Essentially, the concept of boundary obstruction explains why some HOTI models exhibit topologically robust lower-dimensional boundary features without being associated with a bulk invariant, which goes beyond standard symmetry-protected band topology. Such unique bulk-boundary physics of BOTIs can be understood in terms of Wannier spectra and real-space symmetry representations. Although the early quadrupole and octupole models [5,6] are argued to be BOTIs now [39], convincing experiments for identifying such a significant new concept of topological band theory are still elusive. Specifically, the existed studies focus on the demonstration of the higher-order topological signatures in the lower-dimension boundary (i.e., the presence of corner states) [14-16,18,30-32]. However, a complete and conclusive characterization of the boundary-obstructed band topology also demands for revealing the closing and reopening of the higher-dimension boundary states, during which the system's symmetry preserves and the bulk energy spectrum remains gapped. In fact, the latter serves as the key observable distinction of the BOTIs from the conventional TIs. Moreover, to the best of our knowledge, so far there is no experimental realization of three-dimensional (3D) surface-obstructed TIs (SOTIs), which feature closed surface band gap at the phase transition point.

Based on a simple $C_{2h}$ model, here we construct a 3D acoustic SOTI and report a complete experimental characterization for such new topological phases [39]. The model is started with an extended 2D Su-Schrieffer-Heeger (SSH) model, and layer-stacked by dimerized interlayer couplings and alternatively flipped in-plane couplings. In order to confirm the intriguing surface-obstructed band topology, we fabricate three representative acoustic samples with identical in-plane structures but different interlayer couplings, which correspond to trivial, transition-point, and topological insulators, respectively. From the measured bulk and boundary responses to local acoustic excitations, we observe 3D gapped bulk states, 2D gapped surface states, and middle-gap hinge states simultaneously in our SOTI sample, in contrast to the trivial sample that exhibits only gapped bulk and surface signals. As for the transition-point case, our measured bulk and boundary spectra manifest clearly a closure of the surface gap, apart from the still gapped bulk states. The never closed bulk gap, the closed and reopened surface gap, and the emergence of middle-gap hinge states, convincingly evidence the occurrence of the surface-obstructed topological transition and consequent bulk-boundary physics. All experimental results reproduce well our full-wave simulations performed with COMSOL Multiphysics (an established commercial soft solver based on finite-element method).

*General considerations and tight-binding model.*—We present first a general consideration for constructing 3D BOTIs, which capture an obstruction that does not exist on periodic systems but does exist on geometries with symmetric open boundaries. A 3D Hamiltonian $H_{3D}(k_x, k_y, k_z)$ with boundary obstruction can be constructed from a 2D Hamiltonian $H_{2D}(k_x, k_y)$ with nontrivial edge or corner signatures [39]. We consider one pair of such 2D Hamiltonians with opposite in-plane couplings so that their sum is topologically trivial. Next, we stack these pairs via dimerized intra- and inter-cell couplings in the $z$-direction, $\gamma_z$ and $\lambda_z$. This arrangement guarantees that in the limit of $\gamma_z = 0$ (but $\lambda_z \neq 0$), there is a single isolated copy of $H_{2D}$ or $-H_{2D}$ on each $xy$-surface of the open-boundary system. The type of boundary obstruction in $H_{3D}$ depends on the obstruction of $H_{2D}$, as well as the dimension of its nontrivial boundary signature. More specifically, as sketched in Fig. 1(a), here we assume that $H_{2D}$ has a bulk obstruction and features fractional charges localized at the $y$-directed edges. In this case, $H_{3D}$ will exhibit surface obstructions on the $xy$- and $yz$-surfaces. The former can be understood by considering the limit of $\gamma_z = 0$, where there is a single isolated copy of $\pm H_{2D}$ on each $xy$-surface, and the latter can be understood by adding 2D symmetry-protected topological states to the $yz$-surfaces to cancel the surface charges [39]. Similarly, if $H_{2D}$ has an edge obstruction and features fractional charges at its corners, the layer-stacked 3D open system will exhibit an obstruction at hinges.

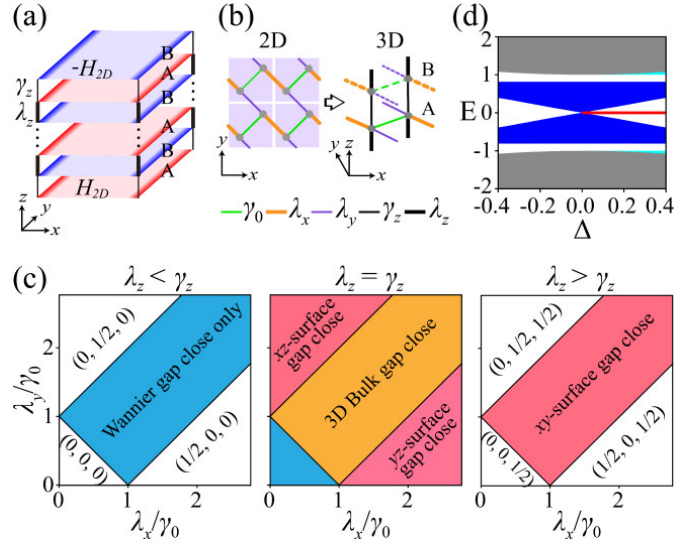

FIG. 1. Tight-binding model. (a) Schematic illustration for realizing a SOTI. (b) Extended 2D SSH model and our AB-stacked 3D unit cell. The layers A and B carry opposite in-plane couplings, as indicated by the solid and dashed lines. (c) 3D phase diagram plotted with three typical slices. (d) Energy spectra simulated for finite lattices with fixed in-plane couplings but different interlayer couplings characterized by $\Delta = \lambda_z - \gamma_z$. The phase transition happens exactly at $\Delta = 0$, featuring clearly a gap closure of the $yz$-surface states (blue) and the emergence of the $y$-directed hinge states (red).

In this Letter, we focus on the surface-obstructed band topology. As shown in Fig. 1(b), we start with an extended 2D SSH model, whose Hamiltonian reads $H_{2D}(k_x, k_y) = (\lambda_0 + \lambda_x \cos k_x + \lambda_y \cos k_y)\sigma_1 + (\lambda_x \sin k_x + \lambda_y \sin k_y)\sigma_2$. Here $\lambda_0$ is the intra-cell hopping, $\lambda_x$ ($\lambda_y$) is the inter-cell hopping along the $x$ ($y$) direction, and $\sigma_{1,2}$ represent Pauli matrices. This model features 1D zero-energy edge states along the $x$ or $y$ direction in the case of $\lambda_y > \gamma_0 + \lambda_x$ or $\lambda_x > \gamma_0 + \lambda_y$ (see *Supplemental Materials*). Once introducing dimerized interlayer couplings, we obtain a 3D Hamiltonian $H_{3D}(k_x, k_y, k_z) = \rho_3 H_{2D}(k_x, k_y) + \lambda_z \sin k_z \rho_2 \mathbb{I} + (\gamma_z + \lambda_z \cos k_z)\rho_1 \mathbb{I}$, where $\rho_{1,2,3}$ are Pauli matrices acting on the two layer sublattices, and $\mathbb{I}$ is the identity matrix of the *x-y* plane. In particular, the Pauli matrix $\rho_3$ endows the alternatively flipped in-plane couplings. Since the three terms in $H_{3D}$ anti-commute with each other, the system is gapless if each term vanishes separately, i.e. only if $H_{2D}$ is gapless and $\lambda_z = \gamma_z$. Figure 1(c) shows the phase diagram for our 3D model, where the phase boundary in each slice is inherited from the 2D one. In addition to the colored regions featuring closed Wannier spectra, there are six bulk-gapped phases distinguished with the topological indices ($P_{z,x}, P_{z,y}, P_{x,z}$) defined by nested-Wilson-loops (see *Supplemental Materials*). Topological phase transition happens only in the cases of closing bulk or surface bands [39]. In particular, the insulator with topological indices (0,1/2,1/2) or (1/2,0,1/2) has a surface obstruction characterized by a quantized hinge charge of 1/2 per unit cell on the $x$- or $y$-directed hinges. The other bulk gapped phases are topologically trivial since they can be adiabatically evolved to the trivial insulator of topological indices (0,0,0).

To confirm the above surface-obstructed topological physics, we calculated the energy spectra for finite lattices with fixed in-plane couplings ($\lambda_x = 2.5$, $\lambda_y = 0.5$, and $\gamma_0 = 1$) but varied interlayer couplings ($\lambda_z = 0.4 + \Delta/2$ and $\gamma_z = 0.4 - \Delta/2$). Figure 1(d) presents the energy spectra as a function of the $\Delta$ value. It shows that as the adiabatic growth of $\Delta$, the bulk states (gray) remain gaped, the $yz$-surface states (blue) close and reopen, and the $y$-directed hinge states (red) emerge as $\Delta = \lambda_z - \gamma_z > 0$. (The hinge states, which are not tied to any nontrivial bulk invariants but associated to nontrivial Wannier-band polarizations, are pining at zero-energy owing to the chiral symmetry and will preserve unless the symmetry is severely broken.) All these facts serve as landmarks of the surface-obstructed phase transition along this path. Note that the surface states emerging on the $xy$ surfaces (cyan, only at $\Delta > 0$) are hardly distinguishable from the bulk states, which are not emphasized here. Surface-obstructed phase transition associated to the closure of $xy$- or $xz$-surface gap can be observed along other paths (see *Supplemental Materials*). The termination dependent obstruction and phase transition serves as a distinctive feature for such newly emergent band topology.

*Acoustic realizations.*—The tight-binding models can be directly implemented with cavity-tube structures in acoustic systems. As sketched in Fig. 2(a), each unit cell consists of four identical air-filled cavities coupled with narrow tubes. Physically, the cavity resonators emulate atomic orbitals and the narrow tubes introduce hoppings between them [24,25,30-32]. To unveil the bulk and boundary physics inherent in the surface-obstructed phase transition, we consider three typical samples with fixed in-plane geometries but varied out-of-plane coupling tubes, which correspond to the topologically trivial (sample 1), transition-point (sample 2), and nontrivial insulators (sample 3), respectively. The structures (see details in Supplemental Materials) provide effectively the onsite energy ~4.31 kHz, the in-plane couplings $\gamma_0 = \lambda_x \approx$ 21.6 Hz and $\lambda_y \approx$ 184.5 Hz, and the out-of-plane couplings: $\gamma_z \approx$ 89 Hz and $\lambda_z \approx$ 33 Hz for the sample 1, $\gamma_z = \lambda_z \approx$ 61 Hz for the sample 2, and $\gamma_z \approx$ 33 Hz and $\lambda_z \approx$ 89 Hz for the sample 3, respectively.

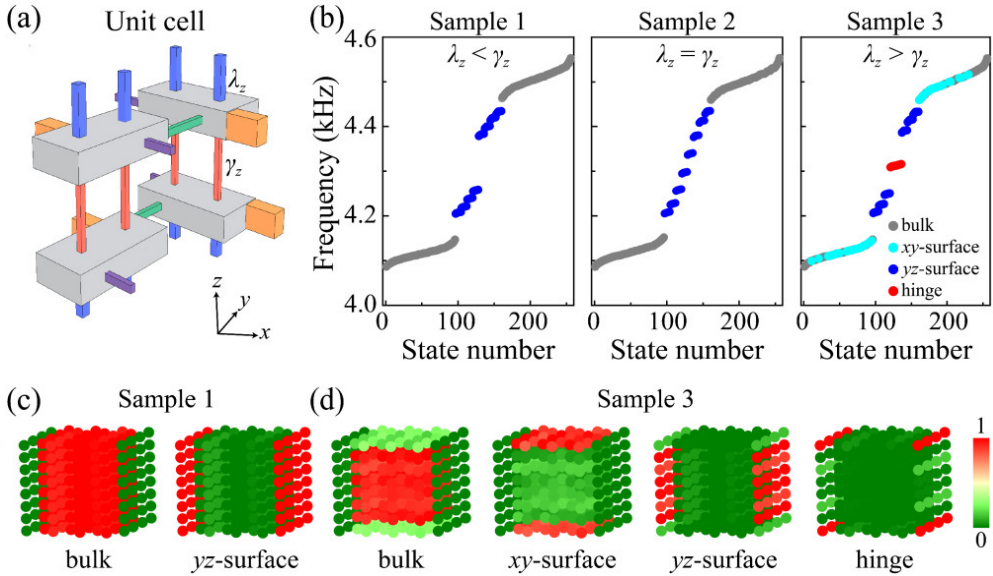

FIG. 2. Full-wave simulations for our acoustic crystals. (a) 3D unit cell made of air cavities (gray) and narrow tubes (color). The sizes of the vertical tubes are tuned to realize topological phase transition, and the in-plane couplings are flipped by the different connectivity between the cavities. (b) Eigenfrequency spectra simulated for three typical acoustic samples of finite sizes. (c) and (d): Averaged intensity distributions of the bulk and boundary states extracted for samples 1 and 3, respectively.

Figure 2(b) provide the eigenfrequency spectra simulated for the above three samples of finite lattices ($4\times 4\times 4$ unit cells). It is observed that for the sample 1 (trivial case) only gapped $yz$-surface states (blue circles) emerge inside the bulk gap, which get closed in the sample 2 (critical case). For the sample 3 (nontrivial case), the $yz$-surface gap reopens and remarkably, hinge states appear in the middle of the $yz$-surface and bulk gaps. Note that the gapped $xy$-surface states (associated to the nonzero $P_{x,z}$), mixing energetically with the bulk states, are distinguished by the eigenstates in real space. All results are consistent with the tight-binding predictions in Fig. 1(d). The gapped bulk in all samples, the closed $yz$-surface states in the sample 2, and the reopened surface gap and newly emerged hinge states in the sample 3, evidence conclusively the occurrence of surface-obstructed topological phase transition and the successful construction of an acoustic SOTI. Figures 2(c) and 2(d) present the field distributions of the bulk and boundary states counted for the samples 1 and 3, respectively. As expected, in addition to the bulk states, the sample 1 hosts surface states localized exactly on the two $yz$-surfaces. A similar result can be found in the sample 2. As for the sample 3, extra states emerge at the two $xy$-surfaces and the four $y$-directed hinges. Interestingly, all the different states are distinguishable in real-space, and in particular, some of the surfaces (e.g., the $xz$-surfaces in all samples) carry the information of bulk states. These facts will facilitate our experimental measurements below.

*Experimental results.*—The band topologies of the above three systems were identified by our acoustic experiments. All samples were 3D-printed with photosensitive resin. Figure 3(a) exemplifies a photo for the sample 3, i.e., the acoustic SOTI with thinner tubes introduced for the intra-cell couplings in the $z$ direction (see inset). To measure the site-resolved local response, each cavity was perforated with two small holes for inserting acoustic source and detector, which were blocked when not in use. Both the input and output signals were recorded and frequency-resolved with a multi-analyzer system (B&K Type 4182). For simplicity, surface measurements were performed to reflect all the information of the bulk, surface, and hinge states (if any), as suggested in Figs. 2(c)-2(d). To do this, as sketched in Fig. 3(b), the sample surfaces were divided into non-overlapping spatial domains to extract the information of different states (see *Supplemental Materials*).

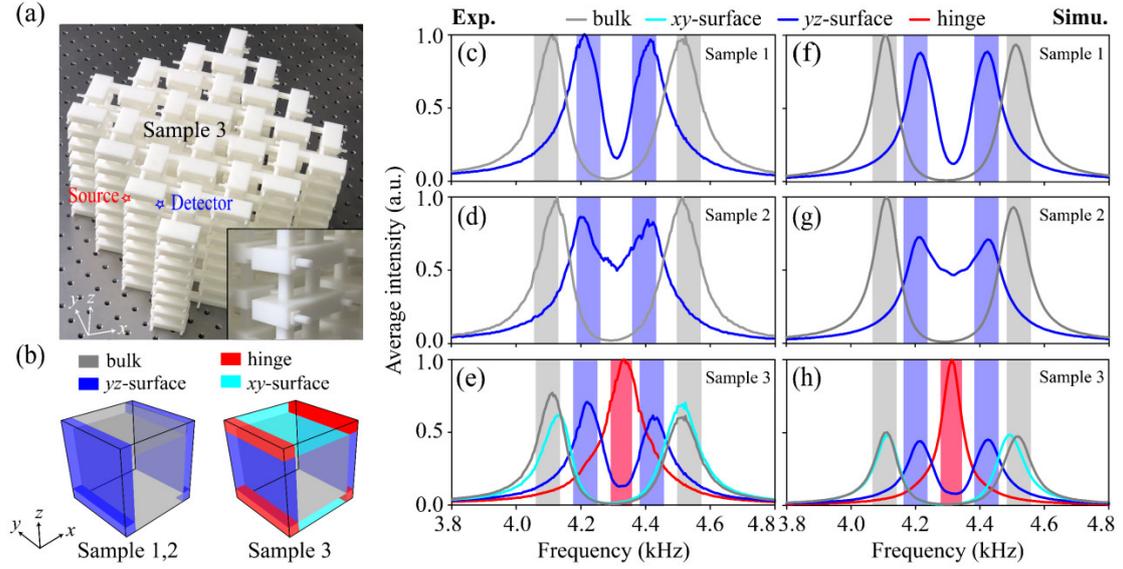

FIG. 3. Experimental measurements. (a) Photograph exemplified for the sample 3. The red and blue stars indicate the locations of the sound source and detector, respectively. Inset: a local view of the sample, which displays the $z$-directed coupling tubes more clearly. (b) Surface domains of the samples divided for calculating average intensity spectra of different states. (c)-(e): Average intensity spectra measured for the bulk, surface, and hinge states (if any) in the three samples. Shaded regions denote the frequency ranges counted for the spatial intensity distributions in Fig. 4. (f)-(h): Numerical comparisons for (c)-(e). All experimental data reproduce well the simulation results.

Figures 3(c)-3(e) present the average intensity spectra of different spatial domains extracted for the three samples. As expected, all the bulk spectra (gray lines) exhibit two intensity peaks (centered at ~4.12 kHz and ~4.51 kHz), which correspond to the bulk bands predicted in Figs. 2(b). The low intensities in between confirm the presence of a wide bulk gap in all the three samples. This fact was further confirmed by our bulk transmission measurements (see *Supplemental Materials*). The $yz$-surface spectra (blue lines) for the samples 1 and 3 exhibit clearly two well-separated peaks (centered at ~4.22 kHz and ~4.43 kHz) within the bulk gap, which identify the gapped surface states in these two systems. In contrast, the surface spectrum for the sample 2 shows high values over the bulk gap. This captures the closure of surface bands in the transition-point system (which serve as the key manifestation distinct from the conventional TIs). The moderate intensity dip around

4.32 kHz arises from the low density of states in the middle of the surface bands (see *Supplemental Materials*). Moreover, the additional hinge spectrum in Fig. 3(e) (red line) shows one prominent peak (centered at 4.32 kHz) inside the common gap of the surface and bulk states, which serves as unambiguous evidence for the existence of topological hinge states. The never closed bulk gaps, the closed and reopened surface gaps, and the emergence of middle-gap hinge states, convincingly evidence the occurrence of surface-obstructed topological phase transition and associated bulk-boundary physics. All the experimentally measured spectra capture well the numerical results presented in Figs. 3(f)-3(h).

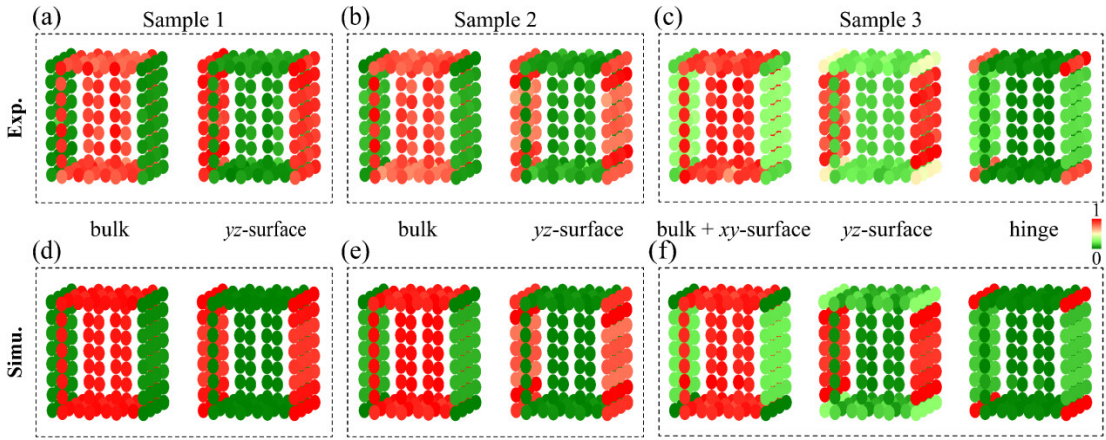

FIG. 4. Field distributions over the sample surfaces. (a)-(c) Intensity patterns measured for different types of states exhibited in the three samples, respectively, integrated over the corresponding frequency ranges shaded in Figs. 3(c)-3(e). (d)-(f) Comparative simulation results.

To further characterize the bulk, surface, and hinge states (if any) of the three samples, we have integrated the pressure intensities over several typical frequency ranges for each surface site individually. Figure 4(a) shows the spatial distributions of the integrated intensity for the sample 1, performed separately for the gray- and blue-shaded regions in Fig. 3(c). The experimental data exhibit clear spatial characteristics of the bulk and surface states as predicted in Fig. 2(c). Similar results are demonstrated in Fig. 4(b) for the sample 2. For the sample 3, in additional to the bulk and $yz$-surface states, the integral performed for the red-colored frequency region in Fig. 3(e) shows a clear hinge-localized pressure intensity [Fig. 4(c)], although the bulk and $xy$-surface states cannot be precisely distinguished in frequency. This is an experimental hallmark of the higher-order band topology for the sample 3. All the experimental field distributions are highly consistent with the numerical results presented in Figs. 4(d)-4(f).

*Conclusions.*—Following a simple 3D model that describes a surface obstruction between distinct topological phases, we designed and fabricated three acoustic samples, which correspond to a SOTI, a trivial insulator, and a transition-point insulator between them. Our measured bulk and boundary responses not only demonstrated the emergence of hinge states, but also identified the essential surface gap closure at the phase transition point, enabling conclusive experimental evidence for such a new concept (i.e., boundary-obstruction) of topological band theory. Excellent agreements were achieved among the tight-binding model, full-wave simulations, and our acoustic experiments. In contrast to those reported previously [33-37], the hinge states observed here are pinned in the middle of the bulk and surface gaps, and thus are more localized in real space and more robust to disorders. In addition, comparing to those dispersive ones, the dispersionless hinge states (*Supplemental Materials*) can freeze the propagation of sound along the hinge. Potential applications can be anticipated for such unique topological hinge states, e.g., acoustic sensing and energy trapping. Last but not the least, our findings can be generalized to electronic, cold atom, photonic, and elastic/mechanical systems.

**Acknowledgements**

This project is supported by the National Natural Science Foundation of China (Grant No. 11890701, 12004287, 12104346), the Young Top-Notch Talent for Ten Thousand Talent Program (2019-2022), and the Fundamental Research Funds for the Central Universities (Grant No. 2042020kf0209).